\documentclass[preprint2]{aastex}

\slugcomment{manuscript for AJ, version 3 150716}

\shorttitle{URAT1 astrometric catalog}
\shortauthors{Zacharias et al.}

\begin{document}


\title{The First U.S.~Naval Observatory Robotic Astrometric 
       Telescope Catalog (URAT1)}


\author{N. Zacharias\altaffilmark{1},
        C. Finch\altaffilmark{1},
        J. Subasavage\altaffilmark{2}, \\
        G. Bredthauer\altaffilmark{3},
        C. Crockett\altaffilmark{2},
        M. Divittorio\altaffilmark{2},
        E. Ferguson\altaffilmark{1},
        F. Harris\altaffilmark{2},
        H. Harris\altaffilmark{2}, \\
        A. Henden\altaffilmark{4},
        C. Kilian\altaffilmark{1},  
        J. Munn\altaffilmark{2},
        T. Rafferty\altaffilmark{5},
        A. Rhodes\altaffilmark{2},
        M. Schultheiss\altaffilmark{2},
        T. Tilleman\altaffilmark{2}, \and
        G. Wieder \altaffilmark{1} }
\email{nz@usno.navy.mil}

\altaffiltext{1}{U.S.~Naval Observatory, Washington DC}
\altaffiltext{2}{Naval Observatory Flagstaff Station, AZ}
\altaffiltext{3}{Semiconductor Technology Associates, San Juan Capistrano, CA} 
\altaffiltext{4}{AAVSO, 49 Bay State Rd., Cambridge, MA 02138}
\altaffiltext{5}{USNO, retired}


\begin{abstract}
URAT1 is an observational, astrometric catalog covering most of the
$\delta \ge -15^{\circ}$ area and a magnitude range of about R = 3 to 18.5.
Accurate positions (typically 10 to 30 mas standard error) are given for 
over 228 million objects at a mean epoch around 2013.5. 
For the over 188 million objects
matched with the 2MASS point source catalog proper motions 
(typically 5 to 7 mas/yr std.~errors) are provided.  These data are 
supplemented by 2MASS and APASS photometry.  Observations, reductions and 
catalog construction are described together with results from external 
data verifications.  The catalog data are served by CDS, Starsbourg (I/329).
\end{abstract}


\keywords{astrometry, proper motions, catalogs}

\section{Introduction}

The U.S.~Naval Observatory (USNO) Robotic Astrometric Telescope (URAT)
project was conceived as next step beyond the successful USNO CCD
Astrograph Catalog (UCAC) project \\
\citep{ucac4} providing accurate 
reference star positions on the sky at current epochs before 
Gaia\footnote{sci.esa.int/gaia, www.cosmos.esa.int/web/gaia} 
data become available.
Original plans called for a new, dedicated 1-meter class telescope
(\cite{z2002}, \cite{z2004}, \cite{laux2005}, \cite{z2006}).
Primarily due to its cost, the instrument was not built.
However, the detector development initiated by a
Small Business Innovation Research (SBIR) through the Office of 
Naval Research lead to the manufacturing of the world's largest, 
monolithic detector, the STA1600.
In 2008 funding became available for a large focal plane instrument
consisting of 4 of those 111 million pixel detectors.
The USNO ``redlens" astrograph \cite{redls} previously used for 
the UCAC survey \citep{ucac4}
was completely rebuilt by the USNO instrument shop in Washington DC.
A single truss-tube structure now joins the old lens and the new
focal plane dewar, utilizing the 9 degree diameter field of view.

After initial tests in September 2011 in Washington DC, the telescope 
was deployed to the Naval Observatory Flagstaff Station (NOFS) in 
Arizona in October 2011.  Upgrades to the dome for lightning protection
were installed, subsequent hardware issues resolved, and operational
robustness improved.  Survey operations began in April 2012.  

A wide dynamic range of stars between about R = 3 and 18.5 mag are 
being observed with a combination of long exposures
and short exposures with an objective grating.
With multiple sky overlaps per year and its high precision URAT is
aiming at 2 specific goals: first, to establish a highly accurate,
dense, deep optical reference frame at current epochs; and second, to
identify nearby stars without selection effects by directly observing
trigonometric parallaxes in an all-sky survey.

Unfortunately the project was delayed due to various technical and
funding issues and is now running close to Gaia data releases.
However, accurate reference stars in the northern hemisphere (URAT1) 
are available now, over a year prior to an anticipated first Gaia 
data release.  
Northern hemisphere observing was completed in June 2015
and the instrument is being packed up for deployment to the 
Southern Hemisphere.
More information about this project and potentially upcoming
data releases is available at our web page 
\url{www.usno.navy.mil/USNO/astrometry/optical-IR-prod/urat} .

The main purpose of URAT1 is to provide the astronomical community
with a reference star catalog for current epochs about 4 times more
precise than UCAC with a density similar to the Two-Micron All Sky
Survey (2MASS) \citep{2mass}, which is on average a factor of 
4 improvement over UCAC4.
Science drivers for accurate positions include:
assist in predictions of solar system bodies occultations,
improve accuracy of positions of moving objects referenced to
the sky background (inertial reference star catalog),
and link of the radio and optical coordinate systems.

Although URAT1 proper motions are very useful for stars fainter
than the UCAC limit, they are only preliminary.  Particularly
for stars in the 8 to 14 mag range UCAC4 has better proper motions
than URAT1 with corresponding more precise positions at much
earlier epochs than 2014.

URAT1 is neither complete (maybe on 90\% level) nor free of contamination 
(expected to be less than 0.5\%).
Systematic errors in positions are estimated to be on the 10 mas level,
while systematic errors in proper motions are on the 1 to 2 mas/yr
level.

The URAT1 catalog is presented as a collection of binary data 
files (total about 18 GB) with index file, auxiliary software and info.
These data are served by CDS, Strasbourg.  There is no data release
on DVD.
This paper accompanies the URAT1 catalog release and provides the
scientific rationale, reduction details, and results.  User of URAT1
are also referred to the ``readme" file included with the data release.


\section{Observations}

\subsection{Instrument}

Basic information about the telescope and camera is provided in
Table 1.  All observations utilize the same 5-element ``redlens"
which also was used for the UCAC program.  A single, fixed bandpass
(680 to 762 nm) is provided by the dewar window which serves as filter.
The bandpass for URAT was pushed as far red as possible to reduce
effects of atmospheric turbulence on the astrometry.

Focusing is performed by moving the lens because the back end
dewar is much heavier than the lens.
The dewar hold time is just over 24 hours, during which time
about 25 kg of $LN_{2}$ are boiled off. 
To compensate for this change in weight, a counterweight is moved
along an arm on the opposite side of the pier to keep the telescope 
balanced.  

A custom made shutter from Bonn University (Klaus Reif) is used.
The layout of the focal plane is shown in Fig.~1.  The gap between
the main CCDs (labeled A to D) is 1200 arcsec.  Each of the 4 main 
CCDs covers a sky area of 2.65$^{\circ}$ by 2.65$^{\circ}$ = 
7 square degrees.

For part of the observing program (see below) a diffraction grating
is mounted in front of the lens.
First order diffraction images saturate for stars
about 5.0 mag brighter than for the central images, thus expanding
the dynamic range of the survey enormously. 
The STA1600 CCDs also feature clocked anti-blooming (CAB) which drains
electrons near saturation to avoid bleeding of charge into neighboring
pixels.  This allows to obtain accurate centroid fits of stars about
1.5 to 2.5 magnitudes brighter than traditional saturation, thus
further extending the dynamic range of URAT data toward bright stars.
Fig.~2 shows the brightness range of stars covered by various
exposure times and modes of operation.  

The STA1600 CCDs are back-side illuminated with optimized quantum 
efficiency near the URAT bandpass.  The CCDs have few to none column 
defects.  The CCDs are operated at normally $-100^{\circ}$C.  
Each main CCD has 8 readout ports going to one side of the CCD.  
All 4 CCDs are read in parallel
in about 20 seconds with a read noise of about 10 e$^{-}$ RMS.  
Full well capacity is at about 80 ke$^{-}$ and the gain is set to
fully utilize the available dynamic range with 16 bit output.

A fiberglass dome with added metal mesh and other lightning protection
measures is used.  To improve the dome seeing, air is sucked out from 
the floor near the telescope mount which provides a steady air flow
down through the dome slit while observing.

\subsection{Survey fields and exposures}

A basic pattern of survey fields was adopted with the same gap size 
between adjacent telescope pointings CCD footprints as the gap size
between the 4 CCDs in the focal plane.
To fill in the gaps, 2 more of the basic pattern of fields were
overlaid with diagonal offsets.  For each of these 3-times basic 
pattern of fields, a 5-fold dither pattern was adopted leading to a
total of 15-fold overlap of fields.  Due to the gaps this results
in an average of 12 fields covering every area on the sky.

A 60 sec and a 240 sec exposure are taken on each such survey field.
To cover a wide parallax factor range, the night is split into 5 parts
of equal duration.  Utilizing 1 dither position per part of the night,
the corresponding 3-times basic pattern of fields is observed
during each part of the night.  Thus the entire pattern of
15 overlap of fields is observed once per year.  Priority is given
to high-declination fields.  The southern limit of the survey is
then determined by the available hours of observing which varies
by season.

The above mentioned regular survey operation is run for about
3 weeks in a month without an objective grating.  During the week 
around full Moon, the same pattern is observed again with short 
exposures (mostly 10 and 30 sec, sometimes 20 and 30 sec) and with 
an objective grating mounted in front of the lens.  

Fig.~3 shows the mean number of observations per star (around 
magnitude 14, i.e.~best case with data from almost any exposure)
as found in the URAT1 catalog.  
More color plots, like mean number of URAT1 stars per square degree
are provided with the data release in the ``info" folder.
URAT1 covers almost all northern sky and most of the area 
$\delta \ge  -15^{\circ}$, plus the far south area around Pluto.

\subsection{Robotic operations}

The entire operations is controlled by a single Linux computer.
The top level URAT operations control software and reduction code
is written in Fortran.  C-code routines are called for hardware
interfaces to 3 systems: telescope, main CCDs, guide CCDs.
Most telescope functions are routed through a Galil controller.
HomeDome provided an interface to control dome shutter and dome 
rotation. 

Every afternoon an ``observer" needs to come to the telescope 
to unhook the $LN_{2}$ line, verify safety of operations, and
check on system status. 
A single command initiates auto-mode operations.  The observer then
notes the time of the $LN_{2}$ fill and paths to save/backup the night's
data.  The counter weight position is then automatically calculated
and moves to position. 
The system is then in a loop mode checking weather conditions and time.
The dome automatically opens around sunset and 
blowers turn on to equalize the temperature in the dome.  
A focus sequence is run and evaluated during twilight and survey 
observations begin when skies are sufficiently dark.
System e-mails are sent for dome open/close and emergency notifications.
Observing progress can be monitored remotely with a command line interface.

The control software has a wait loop before each observation where
weather and safety conditions are monitored.  A Boltwood cloud sensor
unit is used for checking weather conditions and UPS units are
monitored for power outages.  In case of bad weather, the software will
close the dome and wait for the conditions to improve.  In case of
power failure, UPS batteries are sufficient for the software to close
the dome and power off the computers.  A watch script is used to
monitor the control code.  If the software hangs up, the watch script
will initiate a shutdown sequence, closing the dome and parking the
telescope.  The HomeDome system itself will shut down the dome if it
looses communication with the control computer.

A triple fail-safe system is used to prevent the telescope from
hitting something.  Software limits are set for safe operations.
In case they fail, optical limit switches stop the telescope motion.
In case of a complete telescope control failure, mercury limit
switches directly cut power to the telescope motion in case
it exceeds limits in telescope pointing with respect to the horizon. 

At the end of the night the pixel data and log files are ftp'ed
to a second Linux computer and reductions and backups are initiated
automatically, which sometimes run into early next night.
The ``observer" or instrument shop personnel
connects the $LN_{2}$ line sometime in the morning before the auto-fill
begins.  
The only other human interaction is for putting the grating on/off
once a month, checking on desiccant cartridge at the lens, changing 
hard disk drives about once every week or two, and troubleshooting.
The project encountered a fair amount of down time early on due
to dome upgrades for lightning protection and associated side effects.
Recent operations are very stable with only occasional need for a
re-boot of the control computer and/or Galil interface and 
implementing control software fixes after mandatory operating
system kernel updates.

\subsection{Guiding and focus control}

All exposures of 30 sec duration or longer are guided.
The 3 smaller CCDs mounted at the edge of the field of view are
used to take about 4 sec long exposures.  These images are read out
and evaluated in real time.  Mean telescope pointing corrections are
thus derived every about 6 sec and fed to the mount before the
next cycle of guide exposures begins.

The 3 guide CCDs are mounted at different distances with respect
to the focal plane: intra-focal, in focus, extra-focal.
At the end of a guided exposure the mean observed image profile
width of each guide CCD is used to adjust the focus if needed.
The complex lens is sensitive to temperature and temperature
history, so frequent (about 20) focus changes are performed 
throughout the night.

\subsection{Quality control}

After the automatic pixel processing is completed, a project scientist
generates and looks at 5 pages of quality control plots of the data 
of the previous night.  Occasionally manual reductions are run in steps
in case of a failure of a system.  Flags indicating possible problems
are raised automatically and the project scientist has access to
various, detailed log files for troubleshooting if needed.

Limits are set on mean image elongation, image profile width,
saturation and limiting magnitude for an exposure to meet survey
quality.  Mean number of detected objects per readout tab are
compared to identify potential missing data.  Rejected exposures 
are noted and picked up automatically by the scheduling software 
for next night observations.
The quality control pipeline was not fully available
until end of 2012 and early survey data were only spot-checked with
few fields re-observed.

\subsection{Darks, flats, photometry}

Darks for all standard exposure times are taken several times per year.
A few sets of sky flats were obtained throughout the project, taken
near sunset with a piece of white cloth covering the lens.
The pixel-to-pixel sensitivity variations are found to be small
and stable.  Emphasis of this program is on astrometry and only
minimal effort was put into providing URAT bandpass photometry.
Survey quality observations often were obtained during partly
cloudy nights or with cirrus present.  No dedicated observations
of photometric standard stars were performed, and all observations
are taken within about $5^{\circ}$ (20 min) of the meridian.

\subsection{Astrometric calibration observations}

For normal operations the astrograph is on the east side of the pier.
A few times throughout the project, the telescope is flipped to the
west side of the pier allowing observations with the focal plane
rotated by $180^{\circ}$ with respect to the sky.
The same field is observed on the same CCD in both orientations
allowing to detect and calibrate possible magnitude dependent systematic
errors of star positions caused by the instrument alone.  
In another calibration observing run the same fields are being
observed with the objective grating on and using short (typically 30 sec)
and long (240 sec) exposures.  Mapping of $x,y$ centroid positions of
stars observed with such pairs of exposures allows to compare positions
of the same stars as observed with 1st order grating images and
central images, as well as comparing data of saturated and not saturated
images of the same stars.

\subsection{Data used for URAT1}

For the URAT1 catalog reduction 57,129 exposures (of a total of
65,639) from 2012 April 24 until 2014 June 21
were selected, meeting acceptable quality standards.
A total of 14 and 12 exposures of the Pluto field area taken 
2013 Sept 19 and 2014 Sept 06, respectively, were added.
These fields were observed to support occultation predictions.
Altogether data from 380 nights were used for URAT1.
This includes some special observing like bright stars and 
astrometric calibration exposures (see above). 
The distribution of observations by month are shown in Fig.~4.
Clearly a seasonal pattern is seen with fewer exposures than average
taken in winter and summer (monsoon).  Grating survey observing
became routine about half a year into the program.

A small fraction of data was dropped during the reduction process.
In particular, if too few reference stars were matched
or conventional astrometric solutions were poor, those exposures
were just not included in upstream processing.
Similarly all individual ``problem stars" like blended images
were dropped.

The URAT data for this first release was split into 38 sets,
19 each for grating and regular survey data.
All grating observations taken within about a week around each full 
Moon were collected into a set.  All data taken between 2 such 
grating observing runs comprise a set of regular survey data.
In few cases with sparse data, a set extends over about 7 weeks.


\section{Reductions}

The camera output is written to 2-byte unsigned integers in
FITS format with very basic header data, while observing log
files are kept as separate text files.
All raw image processing, photometric and astrometric reductions
were performed with custom Fortran code, building on the software
used for the UCAC program.  IRAF and DS9 were used to display 
sample data and perform image examinations for spot checks,
algorithm development and trouble shooting.
Each of the 4 main camera CCDs was handled separately throughout
the reductions, as if data were taken with a different telescope
for each CCD.

\subsection{Pixel data and centroids}

Raw CCD exposures were bias corrected from overscan data.
A mean dark of corresponding exposure time and epoch was subtracted
and a mean flat applied.  No extra bias frames were utilized.

Background level and noise were calculated from histogram data
of sub-areas.  Objects were detected with a threshold of 4-sigma
above the background.  Basic image profile properties like
elongation and start parameters for the center fit were obtained
from moment analysis of the pixels forming that image.
Object centers were determined from a least-squares fit of the 
pixel data using a 2-dimensional circular symmetric 
Gaussian image profile model function.  Only objects with a successful 
fit were propagated to the next step in the reduction process.

Fig.~5 shows sample results for an exposure taken with the grating.
The CAB feature of the CCD enables a dynamic range of almost 10 magnitudes.
Saturation is at about 30 kADU or instrumental magnitude 6.5.
First order diffraction images have an average elongation of about
1.1 (ratio of major to minor axis of image), and a significantly
larger fit radius than the central images.
The image center fit precision is below 10 milli-pixel over
several magnitudes, including the CAB regime up to 2.5 magnitudes
brighter than the traditional saturation limit.

\subsection{Photometry}

Instrumental magnitudes were derived from the volume of the 
2-dim Gaussian image profile fits, not from aperture photometry.
These magnitudes were calibrated with APASS R and R$-$I data
in a linear fit to derive a zero-point constant between the
instrumental and calibrated URAT bandpass magnitudes.
An example is given in Fig.~6 for a single exposure and CCD.
The formal error of a zero-point constant is typically around 0.01 mag.
Many URAT observations were performed in non-photometric conditions
and URAT photometry is believed to be accurate on a few percent
level at best.
Errors on URAT1 magnitudes as given in the catalog are derived
from scatter of individual observations with 0.01 mag error floor
RMS added.

\subsection{Grating image merge}

For all grating exposures, individual images in the $x,y$ data files
were identified as belonging to a central or first order or higher
order diffraction image by taking into account the brightness,
location and elongation information.  The algorithm to identify
these images was refined by looking at pixel data with the IRAF
``tvmark" feature to highlight various types of images.
An arithmetic mean position was
calculated from a matching pair of first order grating images. 
In addition, all central images were kept as separate observation. 
Thus some stars have 2 observations per exposure based on central
image and first order diffraction images, respectively.
All single first order images and all higher order images are rejected.

\subsection{Astrometric solution}

A subset of the UCAC4 \citep{ucac4} stars, those with astrometry flag
``good" and in the magnitude range of 8.0 to 16.0, were used as reference
stars in a conventional, weighted least-squares adjustment of URAT
data to obtain $\alpha, \delta$ positions.  Using our custom software,
reference stars were matched with the $x,y$ data utilizing telescope
pointing data from the observing log files.

Due to the large field of view, an 8-parameter plate model was adopted
which includes 2 tilt terms ($p, q$) besides the full linear model 
($a$ to $f$, split into orthogonal and non-orthogonal terms):

\[  \xi = a x \ + \ b y \ + \ c \ + \ e x \ + \ f y \ + \ 
          p x^{2} \ + \ q x y  \]
\[ \eta = -b x \ + \ a y \ + \ d \ + \ f x \ - \ e y \ + \ 
          q y^{2} \ + \ p x y  \]

Here $\xi, \eta$ are the standard coordinates (scaled from radian 
to arcsec) and $x,y$ the observed center coordinates of star images 
on the CCD (scaled from pixel unit to arcsec).
This approach is feasible due to the large number of reference stars
available per CCD ranging from a few hundred to several thousand.

The root-square-sum formal errors of UCAC4 reference star 
positions at URAT observing epochs, the formal errors of URAT $x,y$ data 
and an atmospheric turbulence contribution (20 mas for 100 sec exposure, 
scaled by square root of exposure time) were used as weights in the
astrometric solution.  For most stars the largest error contribution
is from the UCAC4 proper motion errors.  Fit results after excluding 
over 3-sigma outliers show on average only about 10\% larger values
than expected from the combination of all known, estimated error 
contributions, which easily could have been underestimated by that 
amount.

Astrometric solution errors are independent of the number of reference
stars available for individual exposures.
There is a significant variation of the astrometric solution error
as a function of Declination, which can be explained by the mean
epoch differences between URAT1 and UCAC4 data. 
The area near the celestial north pole was observed last in the
UCAC4 program leading to the smallest epoch differences to URAT1
data, thus to the smallest UCAC4 position errors at URAT1 epoch
due to error propagation of the UCAC4 proper motion errors.

The CCDs were found to be aligned to $\alpha, \delta$ to within
13 and 36 arcmin, with $x$ being along $\delta$.
The plate tilt parameters were found to be significant and
vary as expected by zenith distance (telescope tube bending
and other effects, like alignment of lens with respect to
focal plane after focus changes).

Fig.~7 shows an example of reference star residuals as a function
of magnitude.  Data from an entire set of grating observations over
5 nights in December 2013 are shown for CCD A.  Systematic position 
differences on the $\pm$10 mas level are seen.  Plots for the other
CCDs look almost identical. 
Remaining systematic errors in UCAC4 alone can explain these differences
\citep{ucac4} and were expected to show up here due to the poor
charge transfer efficiency of the UCAC CCD.
No ``corrections'' to URAT1 as a function of magnitude (over the
non-saturated regime) were applied here in order to not propagate 
such UCAC4 errors into URAT1.

Fig.~8 shows the residuals of the same data set as a function of color.
Most systematic differences are within $\pm$5 mas.
The same type of plots of other data sets look similar.
No corrections for differential color refractions (DCR) were applied.
From theoretical estimates the effect of DCR on URAT1 position data
is found to be below our systematic error floor, which is confirmed
by residual plots as function of color.
Figs.~7 and 8 were generated after correcting for other systematic 
errors as explained next.


\section{Systematic error calibration}

\subsection{Pixel phase error}

Small systematic position errors as a function of the sub-pixel
location of stellar image centers are seen in the astrometric
solution residuals.  As expected these follow a sine-curve as
a function of pixel phase.  The amplitude ($A$) of the sine-curve
is related to $w$, the full width at half maximum (FWHM) of the 
image profile, by an exponential function:

\[  A \ = \ a \ e^{-bw}  \]

The parameters $a$ and $b$ were determined from a linear least-squares
fit to sample data, separately for each CCD and coordinate.
For the smallest observed image profile width in URAT data,
about 1.8 pixel FWHM, the maximal amplitude is about 20 mas. For typical 
data the amplitude is about 10 mas and quickly drops off to insignificant 
levels at FWHM $\ge$ 2.5 pixel.

Individual position corrections to URAT1 observations are then applied 
for each image and coordinate based on the sine-function
of pixel phase value and amplitude as derived for a given exposure's
mean FWHM (from quality control data).
Resulting residuals showed a small remaining effect which was corrected 
by updating the parameters for the pixel phase errors.

\subsection{Field distortion pattern}

Residuals of astrometric solutions using UCAC4 reference stars were
stacked up as a function of $x,y$ coordinate in the focal plane,
separately for each CCD.
Fig.~9 shows an example of such a field distortion pattern (FDP)
for CCD C based on residuals of 140 exposures of acceptable quality of 
a night in March 2014.  The position difference vectors were scaled 
by a factor of 5000.  The FDPs of the 4 CCDs look different due to
the combination of optical distortion (lens and dewar window) and 
individual tilt with respect to the ideal focal plane.  However, 
the level of systematic errors is very similar as given in Table 2.  
 
Sample FDPs of data of the same night but split by magnitude, or
color of stars, or by exposure time were identical within the random 
noise level of about 3 to 5 mas RMS.
Comparing data from different nights, or split by declination 
or early part versus later part of a night displayed typical RMS 
differences of about 5 to 10 mas with somewhat correlated vectors
over the field of view.  This result is not fully understood but
changing temperature gradients across the CCDs are suspected to 
play a role.

The FDPs of several nights were averaged to a mean FDP which was
applied to the $x,y$ data.  After application of the mean FDP the
residual FDP pattern was much smaller but seemed to still show
some systematics. The process was iterated once to arrive at a single
final FDP (per CCD) which was used for calibrating all URAT1 data.

\subsection{General magnitude equations}

Flip observations (section 2.7) provide pairs of observations of the same 
area in the sky and the same CCD with $x, y$ data in 2 orientations
rotated by $180^{\circ}$ between the data sets.
Mapping between 2 such exposures was performed using a full 2nd order
polynomial model (12 parameters) with over 1000 stars in a weighted
least-squares adjustment.  Residuals were then binned and plotted as 
a function of magnitude.

Fig.~10 shows an example for CCD B and exposure pair 7999 versus 8015
(60 sec).  Larger than average scatter is seen at the very bright
end (due to few stars) and the faint end (due to large $x,y$ center
errors).  Note that traditional saturation is at about 6.5 mag on
this scale.  For stars brighter than that, the CAB feature allows for
useful astrometric data.

If there were a pure magnitude equation (systematic position shift
as function of magnitude) caused by the instrument (lens plus camera 
and readout) those systematic errors would show up with doubled
amplitude in these flip observations.  A coma-like magnitude equation
(systematic position shift as a function of the product of magnitude
and $x, y$ coordinate) would show up in the same way.  For UCAC data,
for example, the coma-like systematic errors are in the order of
$\pm$100 milli-pixel (mpx) over 6 magnitudes caused by a poor charge 
transfer efficiency of that CCD.  With URAT1 data we see here nothing
of significance with an upper limit of about 10 mpx / 2 = 5 mpx.

There is a degeneracy between a pure magnitude equation and a coma-like
term as seen in these flip observations.  If they are of exactly the
same amplitude with opposite sign they would cancel out and show a
flat result like seen in Fig.~9, however, that is very unlikely.

Plots for the other CCDs and other pairs of flip observations look
very similar.  There are no indication of a significant magnitude
equation or coma-like term in URAT1 $x, y$ data.

\subsection{Grating images}

Differences between $x,y$ positions of central images and first order
images of grating observations were calculated.  Data from the same 
exposure were analyzed which give only a small magnitude range with 
both 0th and 1st order images having sufficient signal-to-noise (S/N) 
as well as not being saturated.  Similarly $x, y$ transformations
were performed between the 0th order data of 30 sec exposures and 
240 sec exposures of 1st order data to extend the range of overlapping
magnitudes.  In all investigations no systematic difference between the
positions of 0th and 1st order images were found.

\subsection{Astrometric calibration near saturation}

Although the CAB feature of our CCDs allows us to acquire positions 
of stars from successful $x,y$ center fits up to about 2.5 magnitudes 
brighter than saturation, those data are subject to systematic errors 
as compared to unsaturated data.  Regular survey observing provides a 
short (60 sec) and a long (240 sec) exposure of each observed field. 
Thus the long exposure saturates at about 1.5 mag earlier than the 
short exposure.

Using only stars of instrumental magnitude 12 or brighter, the $x,y$
data of a long exposure is matched with the data of the short exposure
using a linear transformation model.  The residuals as a function of
magnitude, separately for each CCD and coordinate, reveal the desired
systematic position offset of saturated data with respect to unsaturated 
data.
An example is given in Fig.~11 for a single night and CCD A.
Data for other CCDs look similar.
Here the instrumental saturation magnitude is near 6.8.  In the range
of 6.8 to 5.3 mag the short exposure is still unsaturated and is
assumed to be ``error free".  The position difference seen in that
range thus shows the systematic error of the long exposure beyond
the traditional saturation.  
For instrumental magnitudes brighter than 5.3 mag the short exposure 
also becomes saturated and begins to show the same relative offset (with 
respect to error free) as the long exposure. 
Assuming the (observed long exposure $-$ true) position difference 
continues on a linear function with magnitude for stars brighter than
(saturation mag - 1.5 mag), we would expect to see a constant
position difference in these plots for stars brighter than about 5.3 mag.
That is what we see for the $x-$coordinate in our example.

For the $y-$coordinate in Fig.~11 we see a drop in the observed position
differences for stars brighter than about instrumental magnitude 5.3
indicating an actual deviation from the simple linear position
correction model.
Empirical corrections were derived based on 2 linear stretches in the
overexposed magnitude regime with corresponding free parameters for the
magnitudes at which each linear stretch begins and ends (i.e.~3 magnitudes
and 2 slope parameters total).  

Data from different nights were analyzed showing a break in the pattern,
some of which correspond to changes in the camera electronics (swap or
replacement of boards which control the readout of the main camera CCDs).
Some data show position offsets exceeding 100 mas at about 2 magnitudes
brighter than traditional saturation, others show total errors of only
about 10 mas.
A total of 6 groups as function of epoch could be identified.
Separate systematic position error corrections for the overexposure 
regime were derived for the data in each of the groups and for each
CCD and coordinate.  The corrections were then checked against the
observed residuals of astrometric solutions using UCAC4 reference stars.
Remaining systematic errors are expected to be $\le$ 10 mas up to about
1.5 magnitudes brighter than traditional saturation and somewhat larger
for even brighter stars.


\section{The Catalog}

Some basic numbers of the URAT1 catalog are given in Table 3.
The URAT1 data are provided by binary zone files each covering 0.2 deg
along declination.  Each entry has 80 bytes of integer data which is
explained in Table 4.   Sample files in ASCII, a detailed ``readme" file, 
index files and basic access code are provided with the public release.
URAT1 is not distributed by USNO, instead is kindly served by CDS, 
Strasbourg through Vizier as catalog I/329.
The distribution of URAT1 stars by magnitude is shown in Fig.~12.

\subsection{Mean positions}

Weighted mean positions were obtained separately for each of the 38
data sets (see above) from the astrometric solutions after having applied 
systematic error corrections for pixel phase error, field distortions, 
and for stars in the CAB regime.
These positions were then combined to a weighted overall mean position
given in the URAT1 catalog.  Thus URAT1 is an observational catalog
providing positions on the International Celestial Reference Frame (ICRF)
via UCAC4 reference stars which are believed to be on the Hipparcos system.
The mean epoch of stars in URAT1 is slightly different for each star 
within the range of 2012.3 to 2014.7.  Most stars have a mean epoch
closer to the center of that range but some objects were observed only
early or late within this range (sky coverage, limiting magnitude).

To make it into the catalog a star needs to have at least 3 observations
or a match with the 2MASS point source catalog within 3 arcsec.
The average number of observations per star is 24.
The distribution of URAT1 position errors (from scatter of individual 
observations) is shown in Fig.~13.

\subsection{Proper motions}

Preliminary proper motions were calculated exclusively from combining 
the observed, mean URAT1 positions with 2MASS positions at about 15 years 
earlier.  Typical proper motion errors are about 5 to 7 mas/yr (see
Fig.~14).  The proper motion errors given in the URAT1 catalog are
formal errors based on the individual epoch difference and positional
errors of the 2 catalogs involved, assuming a constant 80 mas error
for 2MASS data independent of brightness.
Note, stars with few observations (like 1 or 2) and large proper motions
in URAT1 need to be taken with caution, particularly in crowded areas,
because of possible accidental (wrong) matches with 2MASS entries.

\subsection{Added data}

URAT1 observational data are supplemented by 2MASS J,H, and K$_{s}$
magnitudes and some 2MASS flags for about 83\% of stars in common.
The AAVSO Photometric All Sky Survey (APASS) provided B,V,g,r,i magnitudes 
for about 16\% of the URAT1 stars.
The APASS photometry is taken from DR8 plus single observations not
yet published elsewhere.

\subsection{Contamination and completeness}

Possible reasons for false entries in URAT1 include issues with
close doubles/blended images, minor planets, CCD defects,
artifacts near bright stars, and contamination from the 
grating image assignment process.
An upper limit of the number of false objects in URAT1 is estimated 
by the list of 1.1 million objects not found in GSC 2.4 (see below), 
which is less than 0.5\% of the URAT1 catalog entries.

URAT1 is not complete, even in the areas covered by observations.
Blended images, close double stars, and any ``problem" case during
the reductions were just dropped.  The goal here is to provide the
user with an accurate, dense reference star catalog at current epochs.



\section{External Comparisons}

The URAT1 catalog as of November 2014 was extensively validated in-house 
and by selected external reviewers.
Results for positions and proper motions are summarized in Table 5.

\subsection{Primary systems}

At the bright end URAT1 was compared directly to the Hipparcos Catalog
\citep{hip2}.  No significant mean position differences (within about 5 mas)
were found, confirming the URAT1 catalog to be on the Hipparcos / ICRF system.
Observed URAT1 positions of over 66,000 stars in common with the
Hipparcos Catalog were compared at the URAT1 mean observational epoch
using the Hipparcos Catalog positions and proper motions.
For more details see \citep{adela}.

At the faint end URAT1 provided observations of 958 ICRF2 \citep{icrf2}
counterparts.  Excluding outliers, mean URAT1 positions of the remaining
849 sources are consistent with the ICRF within about 10 mas (see Fig.~15).
This is the single, largest systematic error seen with URAT1 so far and
a more detailed investigation including more extragalactic radio sources
is under way.  However, these data indicate an upper limit of a possible
magnitude dependent systematic error in URAT1 over its large range of
over 12 magnitudes (Hipparcos stars to faint end ICRF sources) of about 
10 mas for Dec and less than that for RA.

URAT1 proper motions of the ICRF sources are around zero as expected (Fig.~16).
Similarly over 14,000 extragalactic sources of the second Large Quasar 
Astrometric Catalog (LQAC2) \citep{lqac2} are found in URAT1.  
Their observed mean proper motions as function of declination zone and 
magnitude are typically within $\pm$0.5 mas/yr and up to about 2 mas/yr 
for some samples.  A position comparison of these sources is not meaningful
because of the relative low quality of LQAC2 positions as compared to URAT1. 

\subsection{Deep catalogs}

For the comparison with the PPMXL catalog, 2 zones around RA = $90^{\circ}$
(z1) and 300$^{\circ}$ (z2) going over all declinations and covering the 
10 to 19 magnitude range were selected.  
Comparing PPMXL with ICRF2 large systematic
differences up to about 40 mas are seen, suggesting that some of the
observed URAT1$-$PPMXL differences are caused by errors in the PPMXL.

An extensive comparison to the Sloan Digital Sky Survey (SDSS) data
\citep{sdss} was performed.  
Positions are compared at the URAT1 observational epoch
by using USNO-B proper motions to update SDSS positions.
Overall the systematic differences in proper motions are about 0.4 and 
0.9 mas/yr for RA and Dec, respectively, with variations as a function 
of field on the sky with a standard deviation of 1.8 mas/yr.  
The differences in positions are correlated with the proper motion 
differences and a small magnitude equation is seen
in each coordinate (about $\pm$8 mas between magnitude 14 and 19).

A match of the URAT1 stars with GSC 2.4 was performed (R.~Smart,
private com.~2014).  A total of 1.1 million URAT1 entries are not found 
in the GSC data, which is believed to be very ``clean" and complete.
A spot check of about 30 of those objects (random selection) versus the 
real sky reveals a mix  of different cases.  Some objects are clearly 
seen on the digital sky images, some are in crowded fields, others point 
to very faint objects, no object at all or close to an object, indicating 
a possible large proper motion. 
Thus not all but an unknown fraction of these 1.1 million objects could
be artifacts, or images of moving objects in URAT1.
This result sets a limit of the contamination level in URAT1.
Likely over 99.5\% of URAT1 objects are real, stellar or extragalactic
sources.

\subsection{PM2000}

A detailed comparison of URAT1 data with the PM2000 catalog \citep{pm2000}
was performed (C.~Ducourant \& R.~Teixeira, private com.~2015).
The PM2000 catalog covers the Bordeaux Zone ($+11^{\circ} \le \delta 
\le +18^{\circ}$) of the Astrographic Catalogue (AC) project from 
around 1900.  The AC plates were reduced with Hipparcos and combined
with epoch near 2000 transit circle CCD data to arrive at positions
and proper motions for stars in the about 7 to 16 mag range on the 
ICRF using a global adjustment.  Thus these data are completely 
independent of URAT1 and UCAC4.
URAT1 agrees with PM2000 extremely well for RA with systematic
position differences $\le$ 5 mas over the 8 to 16 mag range and as
function of RA, with no magnitude equation (Fig.~17).
An almost constant position offset of about 10 to 15 mas is seen for
Dec, again with no magnitude equation.  Large position differences
up to 50 mas in Dec are seen between PM2000 and URAT1 at the bright
limit magnitude of PM2000 (V = 7).  This is not seen when comparing
URAT1 with Hipparcos directly.

The systematic proper motion differences between URAT1 and PM2000
are between 0 and 1.2 mas/yr.
A noticeable increase in the RMS proper motion differences is seen
in the galactic plane, which can be explained by confusion of
URAT1 to 2MASS matches and thus contaminating the sample with
some false proper motions.

\subsection{Other comparisons}

Over 57,000 stars of the Lepine and Shara Proper Motion (LSPM) catalog 
\citep{lspm} are found in URAT1.  Due to the respective cut-offs of the 
catalogs only stars with proper motions in the about 150 to 250 mas/yr 
range are seen.  Both the LSPM relative and absolute proper motions were 
compared with URAT1.  There is good agreement for RA but not for Dec.

Vector point diagrams were derived from URAT1 data (R.~Teixeira,
private com.~2015) using 48 near equal area regions on the sphere.
Except for the over-density of proper motions around 200 mas/yr
(again likely a contamination issue in crowded fields) the main 
result is that statistical behavior of URAT1 data is consistent with
predictions of the Galaxy model (Besancon model).
No abnormalities are seen in the spacial distribution of URAT1
proper motions.

Minor planet occultation predictions were analyzed (D.~Harald,
private com.~2014).  Compared with previously used UCAC4 data a 
noticeable improvement for the sample of 90 events observed in 
the past is seen when using URAT1 data.
This check on many random fields indicates good astrometric
performance of URAT1 data for applications which are sensitive
on the 10 mas level of accuracy.


\section{Discussion and Conclusions}

The distribution of URAT1 proper motions on the sky shows the effects
of solar motion and galactic rotation and thus gives some confidence in
the overall accuracy of URAT1 astrometry (no significant spacial biases).
Systematic differences of proper motions between URAT1 and several
other external data are on the 1 to 2 mas/yr level.
At this point it is not clear which data set has the smallest 
systematic errors.

The systematic position error floor of URAT1 data is likely around
10 mas.  URAT1 data match the current celestial reference frame
at the bright end (Hipparcos Catalog) and the faint end (ICRF
counterparts) very well, which suggests insignificantly small
magnitude equations in URAT1 data.
Local, spacial systematic errors of URAT1 positions and proper
motions have not been investigated in detail. However, such
errors are not to be expected due to the ``averaging" URAT observing 
method (same star imaged on different CCDs and different parts of a
CCD) and the homogeneous astrometry of 2MASS.

URAT1 has a very low level of contamination by false entries 
(few tenth of a percent at most) as seen from comparisons with
the GSC and other data.  Possibly some minor planets have made
it into the catalog.  Other false objects will include artifacts
near bright stars or detections associated with blended images.
Some proper motions in URAT1 will be wrong due to mis-matches
with 2MASS, particularly in dense fields.
URAT1 is not intended to be complete but should be complete to
over 90\% in the sky area covered and 3 to 18 mag range due to 
the observing and reduction methods used.

URAT1 can serve as accurate reference star catalog before Gaia
data become available.  The position accuracy of URAT1 is about
4 times higher than for UCAC4 data at its faint end and the
sky density of URAT1 is about 4 times larger than that of UCAC4,
similar to the sky density of 2MASS.



\acknowledgments

The USNO management is thanked for supporting this project:
K.J.Johnston, B.Luzum (the former and current scientific directors of USNO),
R.Gaume, B.Dorland (the former and current head of the astrometry department),
and P.Shankland (the director of NOFS).
Semiconductor Technology Associates (headed by R.~Bredthauer) provided
continued support for the URAT camera and dewar long after delivery.
The American Association of Variable Star Observers (AAVSO) is thanked for 
providing unpublished APASS data for our project.
2MASS was used for near IR photometry and as first epoch of URAT1 
proper motions.
Bill Gray (Project Pluto) is thanked for making available a C code
version of our URAT1 access software:  \url{www.projectpluto.com/urat.htm}
Aladin and Vizier were invaluable tools provided through CDS, Strasbourg.
CDS Strasbourg is also thanked for hosting the URAT1 catalog.
DS9 by the Smithsonian Astrophysical Observatory was used as display tool
  for FITS pixel data files.
NOAO is thanked for IRAF, which was used for image analysis while
  troubleshooting and performing spot checks.
Pgplot by California Institute of Technology was used to produce plots.

\clearpage


\begin{figure}
\includegraphics[angle=0,scale=0.57]{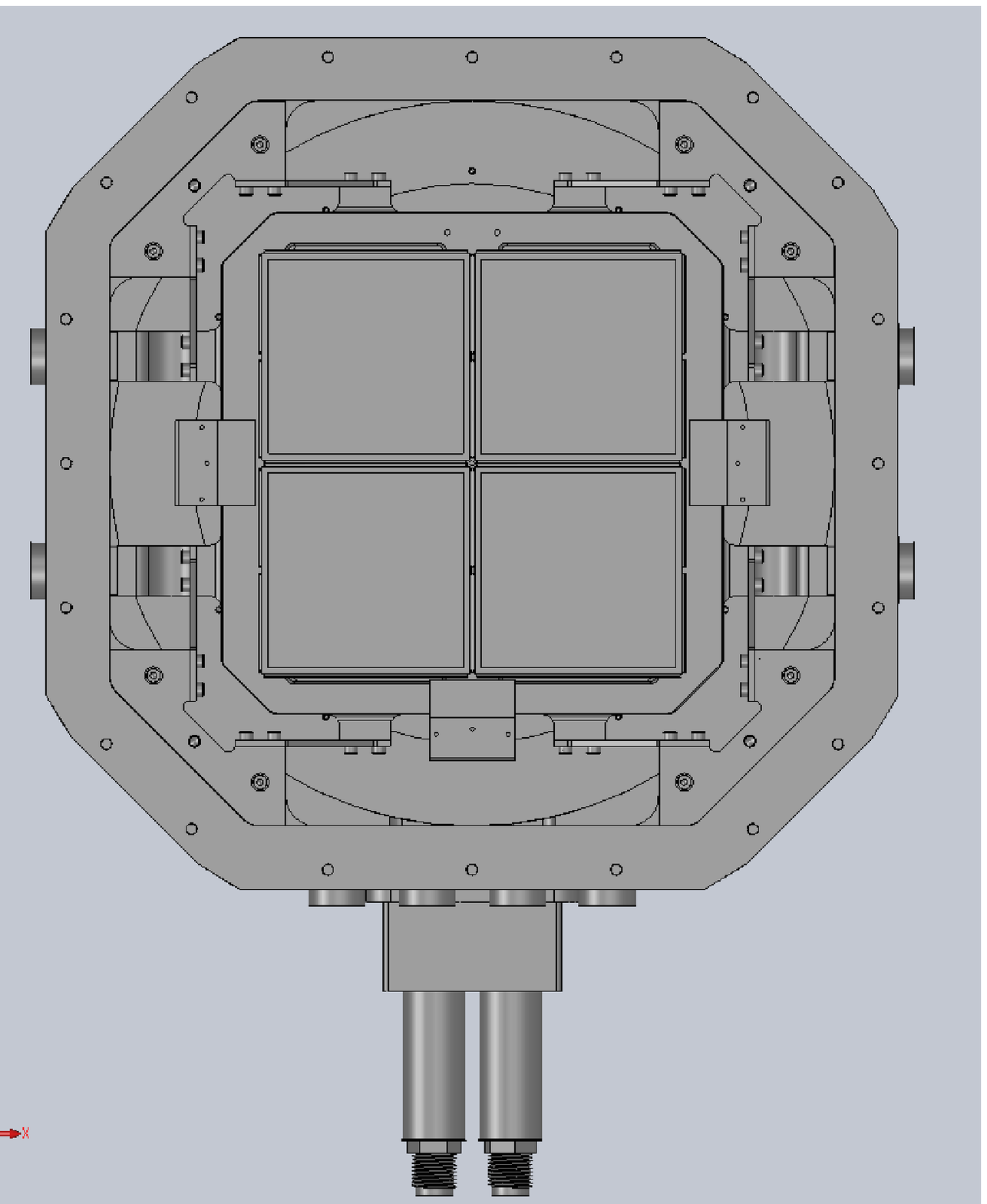}
\caption{Layout of URAT focal plane.  Each of the 4 big CCDs
is 95 by 95 mm in size.  There are 3 more CCDs (right, left, bottom)
for guiding and focus determination.}
\end{figure}

\begin{figure}
\includegraphics[angle=0,scale=0.32]{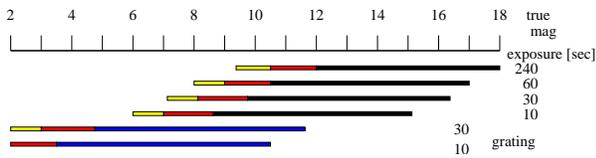}
\caption{Dynamic range of various exposure times and modes of
URAT data. Different shades in color represent from right to left:
unsaturated regime (black and blue), saturated with CAB allowing good 
astrometric results (red), and saturated regime with lower quality 
astrometric calibration (yellow).}
\end{figure}

\begin{figure}
\includegraphics[angle=-90,scale=0.33]{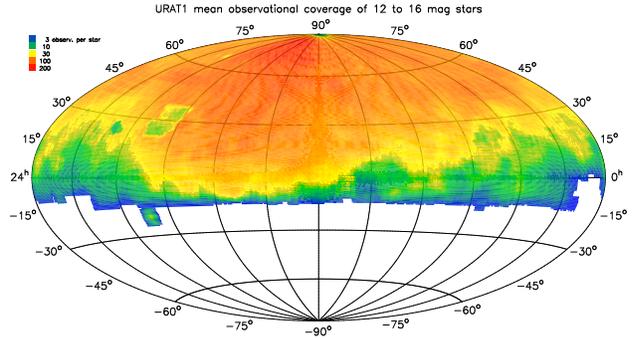}
\caption{Sky coverage of URAT1 catalog data.  Color coding indicates
the average number of observations per star for stars between
magnitude 12 and 16.}
\end{figure}

\begin{figure}
\includegraphics[angle=0,scale=0.47]{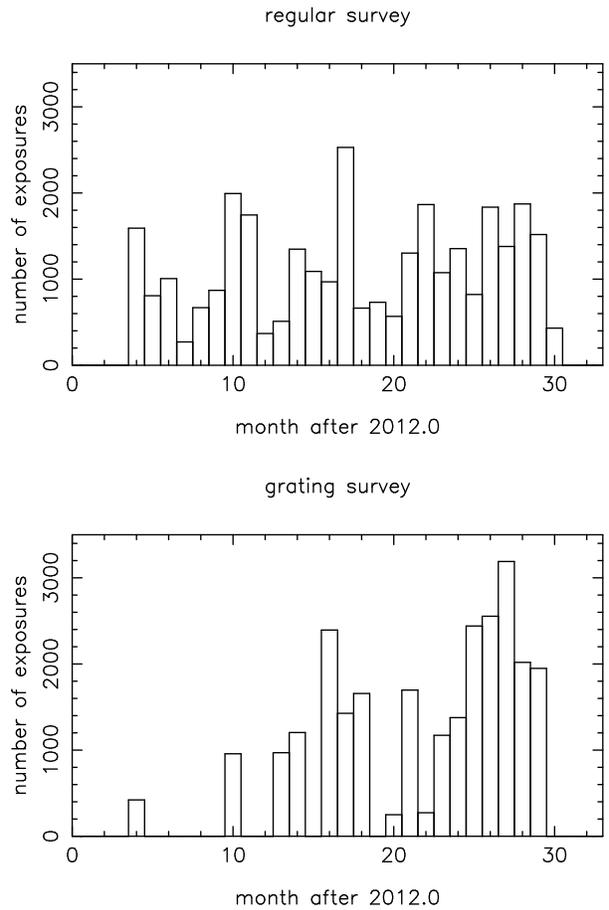}
\caption{Number of exposures used for URAT1 by month after 2012.0
(4 is April 2012, 13 is January 2013).  
Data for the regular and grating survey are shown separately.}
\end{figure}

\clearpage

\begin{figure}
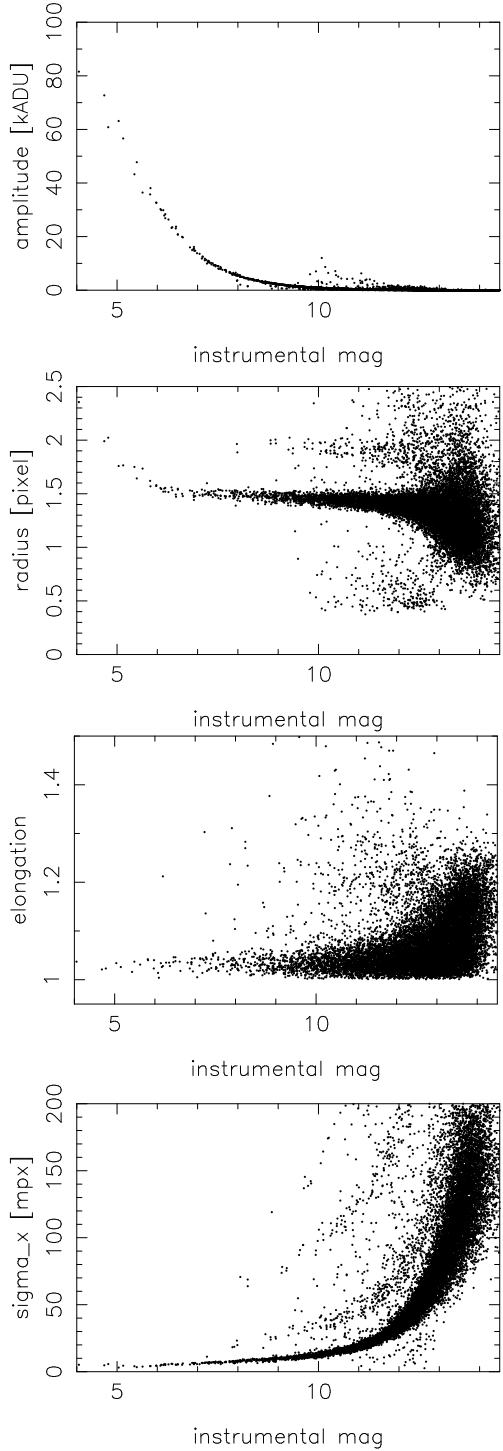

\includegraphics[angle=-90,scale=0.30]{fig05a.ps}
\includegraphics[angle=-90,scale=0.30]{fig05b.ps}
\includegraphics[angle=-90,scale=0.30]{fig05c.ps}
\includegraphics[angle=-90,scale=0.30]{fig05d.ps}
\caption{Individual stellar image profile fit results as a function of
instrumental magnitude; from top to bottom: amplitude, radius, elongation,
and center error per coordinate.  Results for exposure 14064 (10 sec)
of CCD B are shown.}
\end{figure}

\begin{figure}
\includegraphics[angle=-90,scale=0.33]{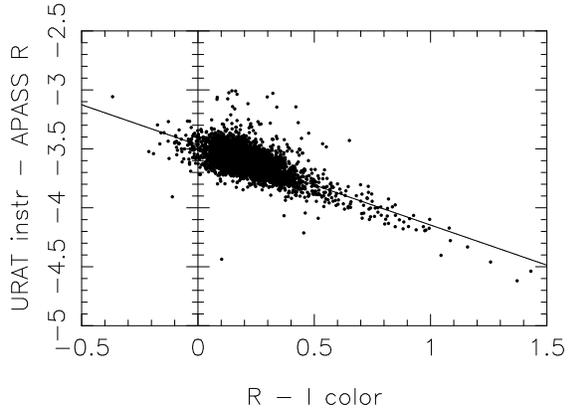}
\caption{Example of color-color data correlation used to
determine photometric constant between URAT instrumental and
calibrated magnitudes per exposure and CCD using APASS R and I data.}
\end{figure}

\begin{figure}
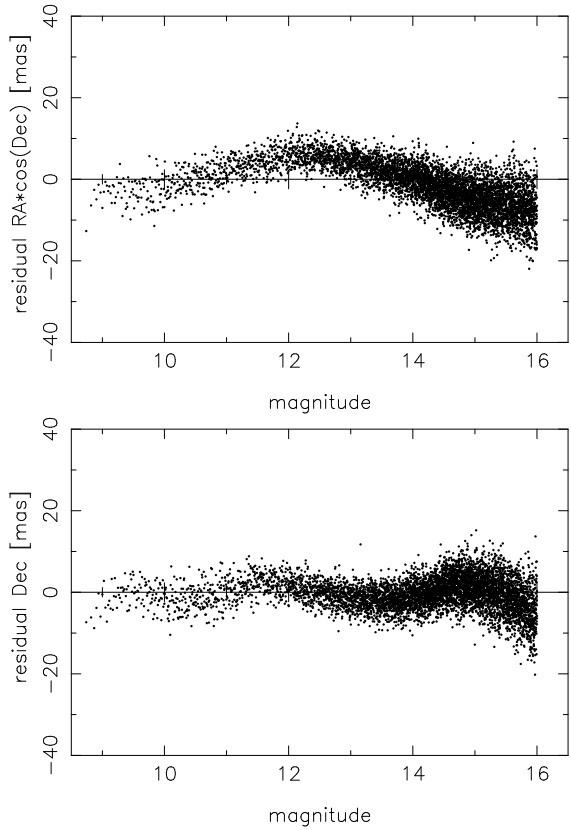

\includegraphics[angle=-90,scale=0.33]{fig07a.ps}
\includegraphics[angle=-90,scale=0.33]{fig07b.ps}
\caption{Example of residuals (URAT1$-$UCAC4) for all grating
observations in set 13 (Julian date nights 6640 to 6644) of CCD A.
Position differences (RA top, Dec bottom) are shown as a function
of UCAC4 magnitude. Each dot is the mean over 1000 observations.}
\end{figure}

\begin{figure}
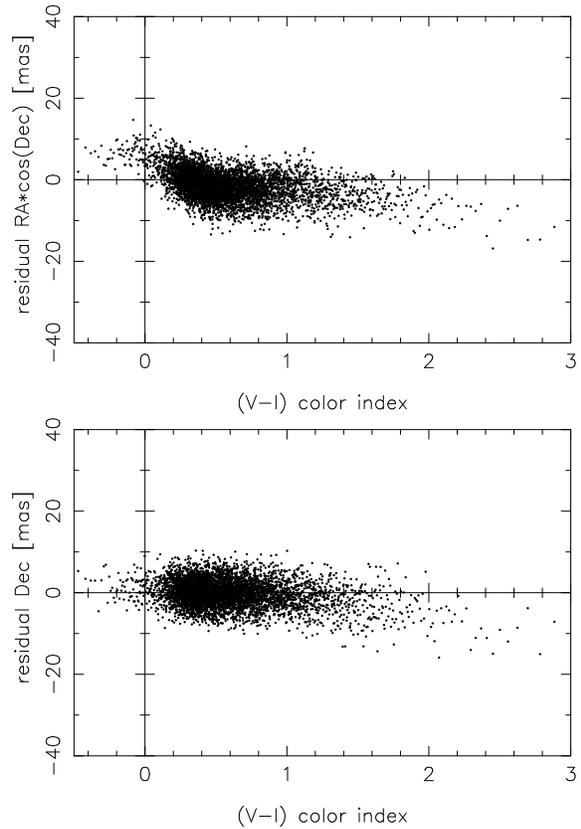

\includegraphics[angle=-90,scale=0.33]{fig08a.ps}
\includegraphics[angle=-90,scale=0.33]{fig08b.ps}
\caption{Same as previous figure for residuals as a function 
of V$-$I color index (from APASS data).} 
\end{figure}

\begin{figure}
\includegraphics[angle=0,scale=0.45]{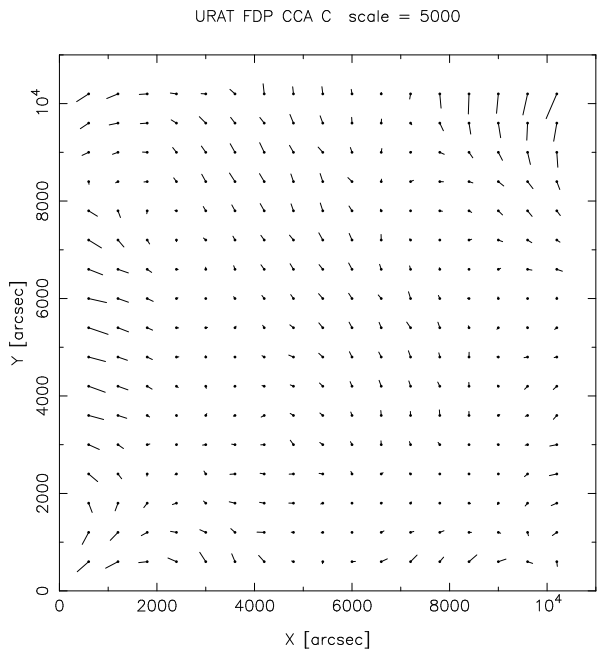}
\caption{Mean field distortion pattern of CCD C as derived from
residuals of a single night. The longest vector is about 80 mas.}
\end{figure}

\begin{figure}
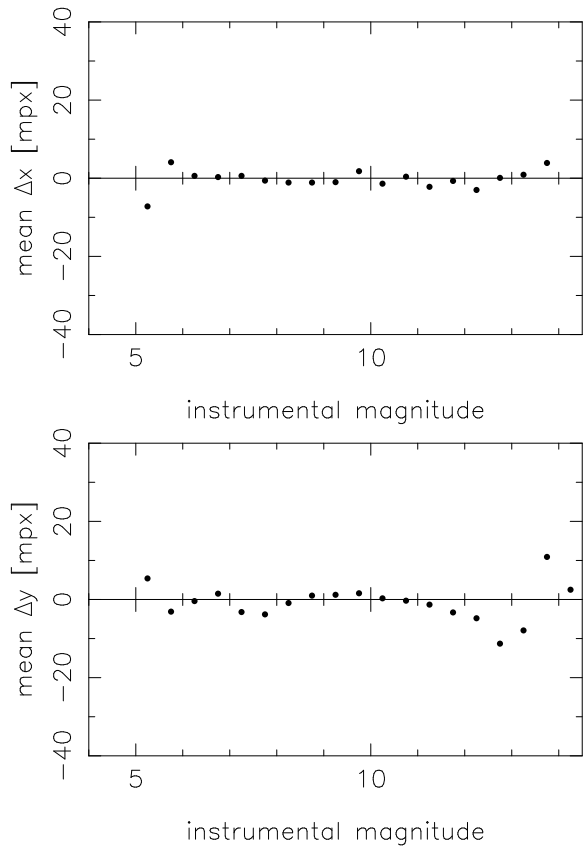

\includegraphics[angle=-90,scale=0.35]{fig10a.ps}
\includegraphics[angle=-90,scale=0.35]{fig10b.ps}
\caption{Position differences of a flip observation (pair of exposures,
regular and $180^{\circ}$ rotated with respect to sky) as a function of
magnitude.}
\end{figure}


\begin{figure}
\includegraphics[angle=0,scale=0.45]{fig11.ps}
\caption{Position differences ($x$ top, $y$ bottom) between 
data from long (240 sec) and short (60 sec) regular survey exposures 
of the same fields.  Results for a single night (JD 6828) are shown
for CCD A.  Each dot represents the mean over 250 stars.
Saturation of the long and short exposure begins at magnitude
6.8 and 5.3, respectively, on this scale. Systematic position
offsets within this range can be attributed to errors caused 
by saturation of the long exposure only.}
\end{figure}

\begin{figure}
\includegraphics[angle=-90,scale=0.33]{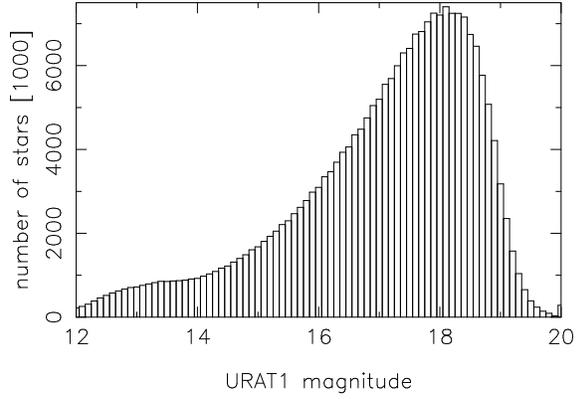}
\caption{Distribution of URAT1 stars by magnitude.}
\end{figure}

\begin{figure}
\includegraphics[angle=-90,scale=0.33]{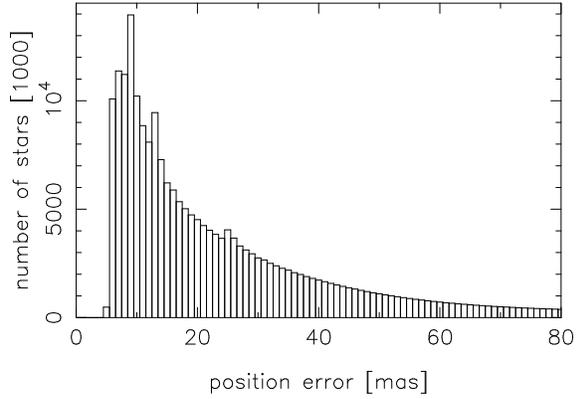}
\caption{Distribution of URAT1 position errors.}
\end{figure}

\begin{figure}
\includegraphics[angle=-90,scale=0.33]{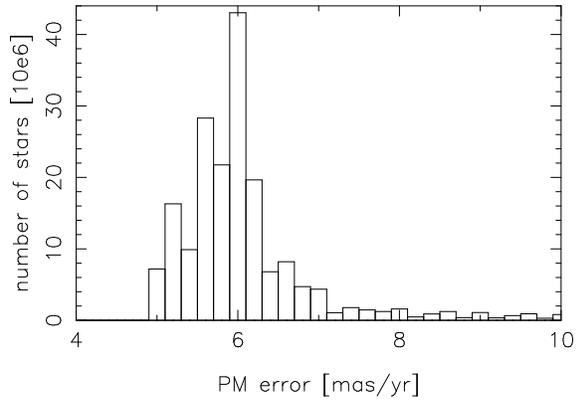}
\caption{Distribution of URAT1 proper motion errors.}
\end{figure}

\begin{figure}
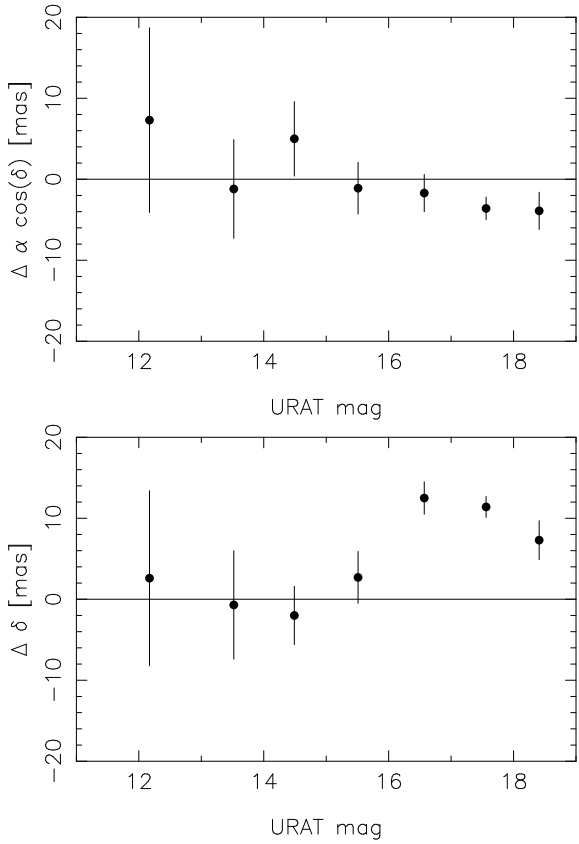

\includegraphics[angle=-90,scale=0.34]{fig15a.ps}
\includegraphics[angle=-90,scale=0.34]{fig15b.ps}
\caption{Weighted mean position differences URAT1$-$ICRF2
of optical counterparts directly observed with the astrograph.}
\end{figure}

\begin{figure}
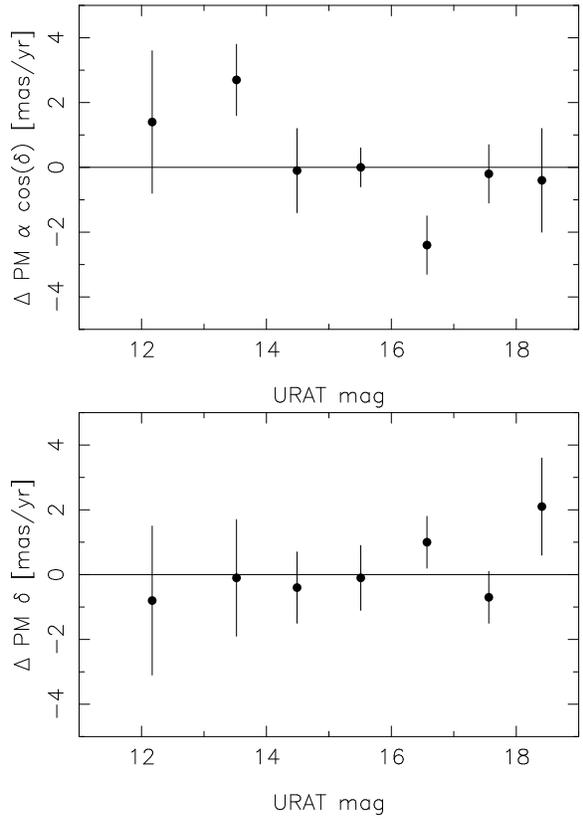

\includegraphics[angle=-90,scale=0.34]{fig16a.ps}
\includegraphics[angle=-90,scale=0.34]{fig16b.ps}
\caption{Mean URAT1 proper motions of the same ICRF2 counterparts
  as used for the previous figure.}
\end{figure}

\begin{figure}
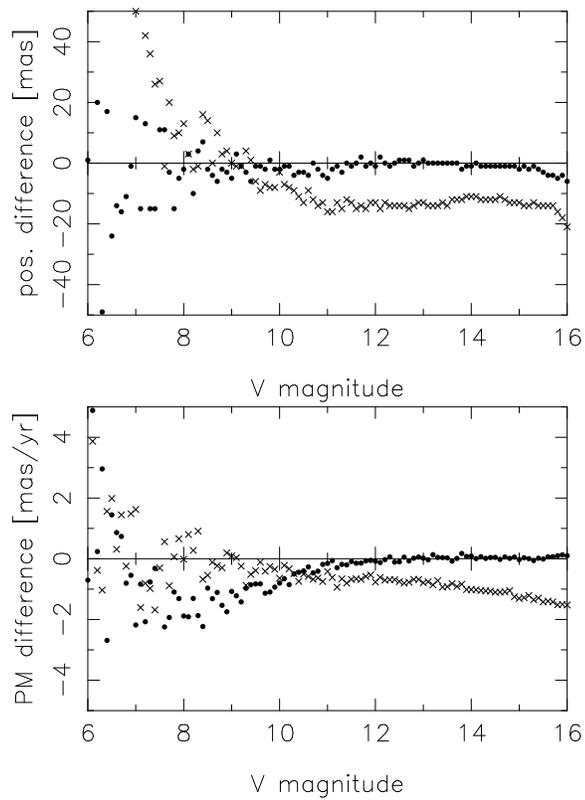

\includegraphics[angle=-90,scale=0.34]{fig17a.ps}
\includegraphics[angle=-90,scale=0.34]{fig17b.ps}
\caption{Mean PM2000$-$URAT1 position (top) and proper motion (bottom)
  differences as a function of V magnitude.  Differences for RA are  
  shown with filled circles, while the crosses represent the differences 
  for the Dec coordinate.}
\end{figure}







\begin{table}
\begin{center}
\caption{Basic data of the telescope, camera, and survey
exposures used for URAT1.} 
\begin{tabular}{llrl}
\tableline\tableline
telescope & astrograph    &         &       \\
          & aperture      &  203    &  mm   \\
          & focal length  & 2060    &  mm   \\
          & bandpass      & 680-762 &  nm   \\
\tableline
camera    & 4 CCDs, each  & 10.5k x 10.5k & px\\
          & pixel size    &    9.0  & $\mu$m \\
          & scale         & 0.905   & "/px\\
          & size, each    & 95 x 95 & mm \\
          & field of view &    28   & sq. deg \\
\tableline
guiding   & 3 CCDs, each  & 2k x 4k & pixels \\
          & scale         &   0.80  & "/px\\
\tableline
regular   & 2 expos./field & 60 \& 240 & sec \\
grating   & 2 expos./field & 10 \& 30 & sec \\
\tableline
\end{tabular}
\end{center}
\end{table}

\begin{table}
\begin{center}
\caption{RMS systematic position offsets (milli-pixel) as a function
  of location in the focal plane (field distortion pattern).}
\begin{tabular}{crrrr}
\tableline\tableline
coordinate& CCD A & CCD B & CCD C & CCD D  \\
          & [mpx] & [mpx] & [mpx] & [mpx]  \\
\tableline
     x    &   22  &   24  &   20  &  15    \\
     y    &   22  &   26  &   22  &  22    \\
\tableline
\end{tabular}
\end{center}
\end{table}

\begin{table}
\begin{center}
\caption{Characteristics of the URAT1 catalog.}
\begin{verbatim}
total numb. URAT1 stars = 228276482
number stars with 1 obs =  10309229
number stars with 2 obs =   8875122
average numb. obs/star  =      24.3

number valid 2MASS data = 188656145  82.64%
no 2MASS match stars    =  39620337    of
stars >=3 obs., no 2MASS=  39079551   URAT1

number valid APASS data =  37010348  16.21%
APASS stars valid B mag =  29313850    of
APASS stars valid V mag =  30057593   URAT1
APASS stars valid g mag =  32340624   stars
APASS stars valid r mag =  32474206
APASS stars valid i mag =  28052917
\end{verbatim}
\end{center}
\end{table}

\clearpage

\begin{table}
\begin{center}
\caption{Description of data items contained in the URAT1 catalog.}
\small
\begin{verbatim}
col item type  unit   description                             
--------------------------------------------------------
  1 ra    I*4  mas    mean RA on ICRF at URAT1 obs.epoch  
  2 spd   I*4  mas    mean South Pole Distance 
  3 sigs  I*2  mas    position error per coord.(scatter)
  4 sigm  I*2  mas    position error per coord.(model)
  5 nst   I*1  --     tot. number of sets star is in      
  6 nsu   I*1  --     n. of sets used for mean position 
  7 epoc  I*2  myr    mean URAT1 obs. epoch - 2000.0           
  8 mmag  I*2  mmag   mean URAT1 model fit magnitude           
  9 sigp  I*2  mmag   URAT1 photometry error                   
 10 nsm   I*1  --     n.of sets used for URAT1 magnitude  
 11 ref   I*1  --     largest reference star flag             
 12 nit   I*2  --     total number images (observations)
 13 niu   I*2  --     n.of images used for mean position
 14 ngt   I*1  --     total n. of 1st order grating obs.
 15 ngu   I*1  --     n. of 1st order grating obs. used
 16 pmr   I*2 0.1mas/yr proper motion RA*cosDec 
 17 pmd   I*2 0.1mas/yr proper motion Dec                       
 18 pme   I*2 0.1mas/yr proper motion error per coord.      
 19 mf2   I*1  --     match flag URAT1 with 2MASS              
 20 mfa   I*1  --     match flag URAT1 with APASS              
 21 id2   I*4  --     2MASS star identification number
 22 jmag  I*2  mmag   2MASS J mag
 23 hmag  I*2  mmag   2MASS H mag
 24 kmag  I*2  mmag   2MASS K mag
 25 ejmag I*2  mmag   error 2MASS J mag
 26 ehmag I*2  mmag   error 2MASS H mag
 27 ekmag I*2  mmag   error 2MASS K mag
 28 iccj  I*1  --     CC flag 2MASS J                         
 29 icch  I*1  --     CC flag 2MASS H
 30 icck  I*1  --     CC flag 2MASS K
 31 phqj  I*1  --     photometry quality flag 2MASS J        
 32 phqh  I*1  --     photometry quality flag 2MASS H
 33 phqk  I*1  --     photometry quality flag 2MASS K
 34 abm   I*2  mmag   APASS B mag                            
 35 avm   I*2  mmag   APASS V mag
 36 agm   I*2  mmag   APASS g mag
 37 arm   I*2  mmag   APASS r mag
 38 aim   I*2  mmag   APASS i mag
 39 ebm   I*2  mmag   error APASS B mag
 40 evm   I*2  mmag   error APASS V mag
 41 egm   I*2  mmag   error APASS g mag
 42 erm   I*2  mmag   error APASS r mag
 43 eim   I*2  mmag   error APASS i mag
 44 ann   I*1  --     APASS numb. of nights                  
 45 ano   I*1  --     APASS numb. of observ.                 
--------------------------------------------------------
\end{verbatim}
\end{center}
\end{table}

\clearpage

\begin{table}
\begin{center}
\caption{Summary of external comparisons of URAT1 catalog data.
  Numbers in parentheses indicate the URAT1 magnitude.}
\small
\begin{verbatim}
positions:
-----------------------------------------------
catalog   dRAcos(D) [mas]      dDec [mas]       
          bright     faint     bright     faint
-----------------------------------------------
ICRF2    +6 (12)   -4 (19)    +3..-2..+12..+7   
Hip.2     0 ( 8)              +5 ( 8)  
PPMXLz1   0 ..+5.. 0 ..+20    0 ..-10..-10..-20 
PPMXLz2  +5..+10..+5 ..+15   +10..-10..-25..-10 
SDSS     +9 (14)   -8 (19)    -6 (14)   +8 (19)
PM2000   +2 ( 9)    0 (15)   +10 (10)  +12 (15)
-----------------------------------------------

proper motions:
----------------------------------------------------
catalog    dRAcos(D) [mas/yr]      dDec [mas/yr]    
           bright       faint      bright      faint
----------------------------------------------------
ICRF2    +2.0 .. -2.0 .. 0.0    -0.5 (12)  +0.5 (18)
LQAC2    +2.2 (12)  -0.2 (19)   -0.5 (14)   0.0 (19)
LSPMa     0.0 ( 8)  -0.5 (18)   +1.0 ( 8)  +4.0 (18)  
LSPMr    -2.0 ( 8)  -2.0 (18)   -7.0 ( 8)  -3.0 (18)  
PPMXLz1   0.0..+0.5..-1.0       -0.8.. 0.0..+1.0     
PPMXLz2  +0.5..+1.0..-0.5        0.0..-0.8..+1.0     
SDSS     +1.0 (14)  -0.1 (19)   -1.7 (14)  -0.2 (19)
PM2000   +1.0 ( 9)   0.0 (16)    0.0 ( 9)  +1.2 (15)  
----------------------------------------------------
\end{verbatim}
\end{center}
\end{table}

\end{document}